%% file: nips_2018.tex
\documentclass{article}

\usepackage[preprint,nonatbib]{nips_2018}

\usepackage{xcolor}
\usepackage{graphicx}
\usepackage[utf8]{inputenc} 
\usepackage[T1]{fontenc}    
\usepackage{hyperref}       
\usepackage{url}            
\usepackage{booktabs}       
\usepackage{amsfonts}       
\usepackage{nicefrac}       
\usepackage{microtype}      

\title{Temporal Unet: Sample Level Human Action Recognition using WiFi}
\author{Fei Wang\textsuperscript{1,3}~~~~Yunpeng Song\textsuperscript{2,3}~~~~Jimuyang Zhang\textsuperscript{3}
~~~~Jinsong Han\textsuperscript{1}
~~~~Dong Huang\textsuperscript{3}
\\
\small{\textsuperscript{1}Zhejiang University~~~~\textsuperscript{2}Xi'an Jiaotong  University~~~~\textsuperscript{3}Carneige Mellon University}\\
\small\texttt{feiwang@cmu.edu, yunpengs@andrew.cmu.edu, zhangjim@andrew.cmu.edu}\\
\small\texttt{hanjinsong@zju.edu.cn, donghuang@cmu.edu}
}

\begin{document}

\maketitle

\begin{abstract}
Human doing actions will result in WiFi distortion, which is widely explored for action recognition, such as the elderly fallen detection, hand sign language recognition, and keystroke estimation. As our best survey, past work recognizes human action by categorizing one complete distortion series into one action, which we term as series-level action recognition. In this paper, we introduce a much more fine-grained and challenging action recognition task into WiFi sensing domain, i.e., sample-level action recognition. In this task, every WiFi distortion sample in the whole series should be categorized into one action, which is a critical technique in precise action localization, continuous action segmentation, and real-time action recognition. To achieve WiFi-based sample-level action recognition, we fully analyze approaches in image-based semantic segmentation as well as in video-based frame-level action recognition, then propose a simple yet efficient deep convolutional neural network, i.e., Temporal Unet. Experimental results show that Temporal Unet achieves this novel task well. 
~\textit{Codes have been made publicly available at}~\url{https://github.com/geekfeiw/WiSLAR}.


\end{abstract}

\input{tex/introduction.tex}

\input{tex/related.tex}
\input{tex/background.tex}

\input{tex/t-unet.tex}
\input{tex/results.tex}

\section{Conclusion}
In this paper we introduce the sample-level action recognition into WiFi sensing community. We fully analysis similar tasks in computer vision, then propose a simple but efficient network, i.e., Temporal Unet, which targets on sample level action recognition by inputting time-serial WiFi distortion induced by human action. Experimental results show advancements of Temporal Unet.

\section*{Acknowledgments}
F.W. and YP.S. are supported by China Scholarship Council.

{\small
\bibliographystyle{IEEEtran}
\bibliography{sample-base}
}

\input{tex/appendix.tex}

\end{document}

%% file: tex/introduction.tex
\section{Introduction}
WiFi devices have been widely studied in human action recognition, such as the elderly fallen detection~\cite{wang2017wifall,wang2017rt}, hand sign language recognition~\cite{li2016wifinger,wang2018csi}, keystroke estimation~\cite{ali2015keystroke,li2016csi}, etc. Compared with cameras, WiFi devices as ubiquitous sensors for human action recognition are more resilient to illumination and occlusion, meanwhile arise less privacy concerns. As our best survey, past work does action recognition by categorizing one complete WiFi distortion series into one action. For this, past work usually (1) applies action detection algorithms to detect the start point and the end point of actions~\cite{wang2017rt,ali2015keystroke}, (2) segments WiFi distortion series by estimated start/end points, (3) and recognizes action based on the segmented series with series-matching algorithms such as the dynamic time warping~\cite{berndt1994using}. From above perspective, past work can be called series-level action recognition, which classifies the whole WiFi distortion series with one single action label.

\begin{figure}[t]
\centering
\includegraphics[width=1\textwidth]{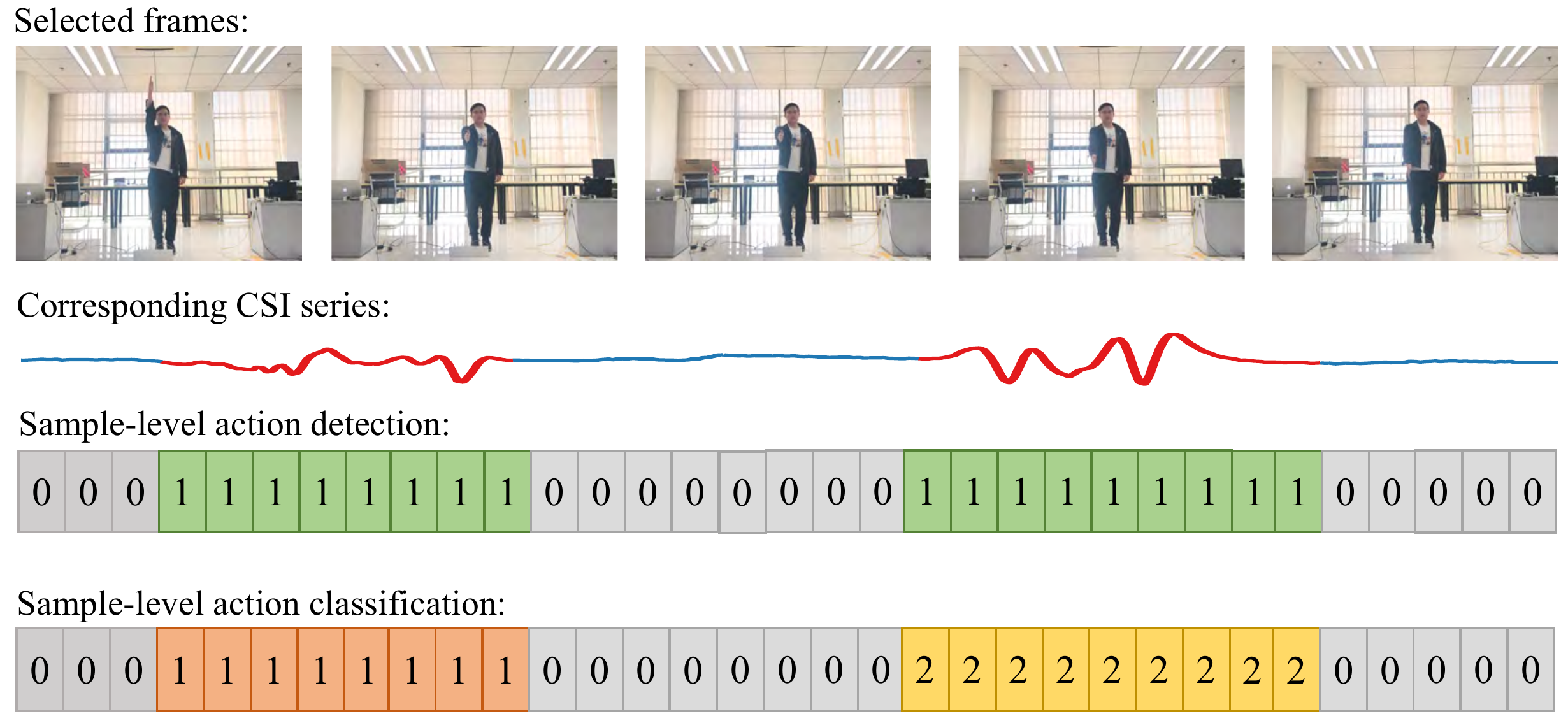}
\caption{WiFi-based action recognition. When one conducts actions~(1st), channel state information~(CSI) of WiFi varies~(2nd). Sample-level action detection is to identify whether one action is conducted or not for every sample on the series~(3rd), and sample-level action classification is further to identify which action is conducted~(4th).}
\label{fig:first}
\end{figure}

In this paper, we introduce a much more fine-grained action recognition task into WiFi sensing domain, i.e., sample-level action recognition, to enlarge the sensing ability of WiFi sensors. In this task, every sample in the whole series should be categorized into one action class, which is critical in
precise action temporal localization, continuous action segmentation, and real-time action recognition.
More precisely, sample-level action recognition roughly falls into the following two tasks. (1) sample-level action detection, and (2) sample-level action classification. As shown in Fig.~\ref{fig:first}~(1st, 2nd), when a user does actions, he distorts WiFi sampling series. The sample-level action detection is to estimate whether a user is doing actions or not for every sampling moment, shown in Fig~\ref{fig:first}~(3rd). Besides, the sample-level action classification also aims to categorize action for every WiFi sample, shown in Fig~\ref{fig:first}~(4th).

However, there are two severe challenges to achieve WiFi-based sample-level action recognition. (1) Since it is hard to identify action only based on one sample, to categorize one sample, we must learn relations from its neighbouring samples. However it is challenging to figure out how many neighbours should be taken into account for sample-level action recognition. (2) The proposed approach should better be unified and applicable for both of aforesaid two tasks, i.e., sample-level action detection and action classification. Nonetheless, these two tasks may require distinct features, which increases the difficulty of approach proposal.

To solve these challenges, we propose the novel Temporal Unet that enables Unet~\cite{ronneberger2015u} ability of learning action features from time-serial WiFi distortion that induced by human actions. Precisely we apply temporal convolutional layers, temporal max pooling layers, and temporal deconvolutional layers along the time axis of sampled CSI series to learn features from the low-level variance and high-level profile. Besides, the stacked temporal layers could cover neighbours of target sample in various time of views~\cite{simonyan2014very}. Further in Temporal Unet, shortcut links between low layers and high layers can combine features in various levels to boost action recognition. With advancements in deep learning design, experiment results show that Temporal Unet achieves good results in WiFi-based sample-level action recognition. Contributions of this paper can summarized as follows.

(1) We introduce WiFi-based sample-level action recognition task into WiFi sensing domain to enlarge WiFi abilities in human sensing. In addition, we fully discuss potential techniques and insights that can achieve this task. 

(2) We propose the novel Temporal Unet for this task, which has been evaluated simple yet efficient both in WiFi-based sample-level action detection and WiFi-based sample-level action classification.

%% file: tex/related.tex
\section{Related Work}\label{sec:related}

\subsection{WiFi-based Action Recognition}
WiFi devices and signals have been explored for human action recognition such as the daily activity recognition~\cite{pu2013whole,wang2014eyes,xi2014electronic,fang2016bodyscan,abdelnasser2015wigest}, health-care usages~\cite{wang2017wifall,wang2017rt,palipana2018falldefi,wang2018csi}, and hand gesture recognition in privacy and security or human-computer interaction applications~\cite{li2016wifinger,wang2017csi,ali2015keystroke,wang2019joint}. Past WiFi-based action recognition work follows 3 main technology roadmaps. (1) Utilizing the statistical values such as the average, variance, and entropy as features of the WiFi time-series to train action classifiers, then applying the trained classifiers to categorize action by inputting the whole WiFi series that correspond to one action~\cite{wang2017wifall,wang2017rt,wang2018continuous,zeng2016wiwho,pu2013whole}. 
(2) Using Dynamic Time Warping~\cite{berndt1994using} to measure distances between the test WiFi series and all training series, then applying K Nearest Neighbors algorithm to predict one action for the test set~\cite{xi2014electronic,fang2016bodyscan,abdelnasser2015wigest,li2016wifinger,wang2017csi,ali2015keystroke}. And to date, (3) designing deep learning approaches such as deep Boltzmann Machine~\cite{salakhutdinov2009deep} and Convolutional Neural Networks~\cite{lecun1998gradient,krizhevsky2012imagenet} as the WiFi-based action classifiers~\cite{wang2017csi,wang2017cifi,ma2018signfi,zhou2018signal,zhang2018enhancement, wang2019joint}. 

Since past work predicts whole WiFi distortion series as one action label, we call it series-level action recognition. In this paper, we do action recognition on every sampled WiFi data, which is much more fine-grained and challenging. To our best survey, this is the first work on the sample-level WiFi-based action recognition.

\subsection{WiFi-based Action Detection}
Our approach can also be used to detect human action, shown in Fig.~\ref{fig:first}, where `1' and `0' represent `doing an action' and `no action', respectively. To this end, our work is related to previous WiFi-based action detection work that detects the start time and the end time of an action. As shown in above studies, action detection usually serves as one part of techniques to segment WiFi series for further series-level action recognition~\cite{ali2015keystroke,wang2017rt,wang2017wifall,wenyuan2018lens,lin2019concurrent}. In addition, past work usually applies threshold-based sliding window in amplitude, variance, entropy, etc., to detect the start time and end time of an action. For example, Wang et al.~\cite{wang2017rt} first computes the mean $\mu$ and the normalized standard deviation $\sigma$ of the non-action state, then uses  $\mu+6\sigma$ as the threshold to determine the non-action and doing-action state for further (fine-grained?) start time and end time detection. Similarly, in~\cite{wang2017wifall}, the local density~\cite{breunig2000lof} of WiFi series in non-action state is used as a threshold to detect human falling state. In~\cite{wenyuan2018lens,lin2019concurrent}, the first-order difference of WiFi series is computed and applied for starting point detection. And in~\cite{ali2015keystroke}, the mean absolute deviation of WiFi series is utilized for a threshold-based for action detection. 

There are two major shortcomings in above approaches. First, thresholds require hyper-parameters, such as the $6$ in $\mu+6\sigma$~\cite{wang2017rt}, which decreases the effectiveness and generalization ability of these methods. Second, above approaches cannot segment a series of continuous actions those have little transferring state between two consecutive actions. Our approach requires no hyper-parameters and no thresholds, whereas it learns boundaries of actions directly from WiFi distortion series, thus overcoming these two shortcomings significantly.

\subsection{Dense Task in Computer Vision}
Unlike generating one action label from the whole WiFi series~(sparse labeling), in our proposed sample-level WiFi sensing task, every sampled WiFi distortion should be categorized to one action. We name it dense WiFi sensing task, which is related to some dense tasks in computer vision domain such as image-based semantic segmentation~\cite{long2015fully,ronneberger2015u,badrinarayanan2017segnet,chen2018deeplab} and video-based frame-level action recognition~\cite{tran2015learning,hara2018can,shou2017cdc,yang2018exploring}. In image-based semantic segmentation, every pixel of the image is labeled as one thing/stuff class such as the sky, giraffe, person, etc., in Fig.~\ref{fig:image}. That is, the output is with the same size, i.e. height and width, as the input image. Similarly, in WiFi-based sample-level action recognition, the output should also be with the same size as the input WiFi series. This similarity inspires us to explore the feasibility of applying deep frameworks that meet the same-size requirement such as FCN~\cite{long2015fully},  U-Net~\cite{ronneberger2015u} and FPN~\cite{lin2017feature} in our proposed task. However it still requires careful designs on the framework targeting to WiFi time series. Besides, our task is related to video-based frame-level action recognition that generates an action label for every frame, shown in Fig.~\ref{fig:video}. The main purpose of this work is to find relations between frames in temporal~\cite{tran2015learning,hara2018can,shou2017cdc,yang2018exploring} as well as the relations between objects including persons in one frame~\cite{gkioxari2018detecting}. 
One main purpose of this category of work is to find action representation from multiple continuous frames ~\cite{tran2015learning,hara2018can,shou2017cdc,yang2018exploring}, which inspires us that learning action representation in temporal would facilitate our  sample-level action recognition task.

In section~\ref{sec:background}, we describe more details on our approach inspiration from previous dense tasks.

%% file: tex/background.tex
\section{Background and Analysis}\label{sec:background}

For simplicity, in this section, we name WiFi-based sample-level action recognition task as WiSLAR. We next analyze relations between WiSLAR and image-based semantic recognition as well as video-based action recognition, show insights from the analysis, and propose our approaches.

\begin{figure}[t]
    \centering
    \includegraphics[width=0.98\textwidth]{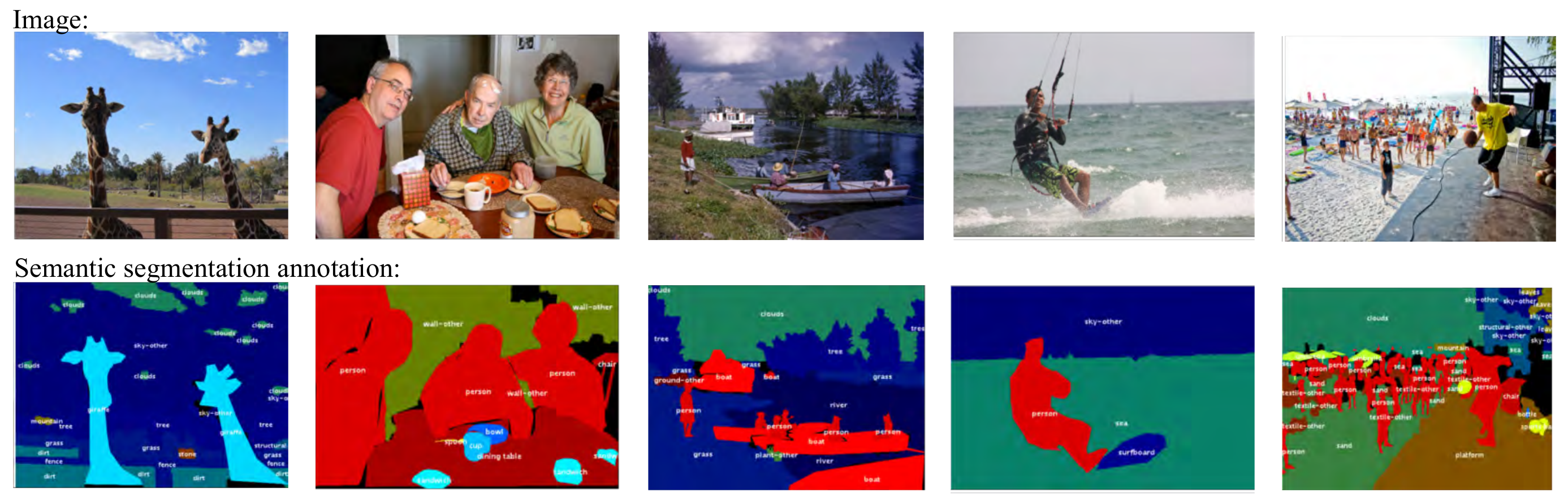}
    \caption{Image-base semantic segmentation samples from COCO-Stuff dataset~\cite{caesar2018coco}. Every pixel in the image is labeled as one thing. }
    \label{fig:image}
\end{figure}

\begin{figure}[t]
    \centering
    \includegraphics[width=1\textwidth]{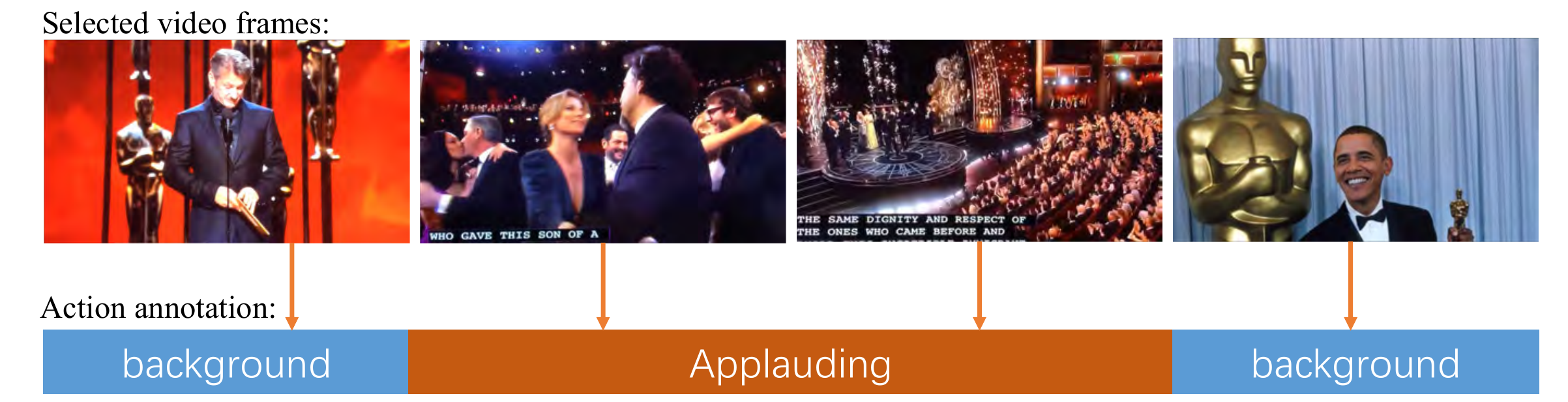}
    \caption{One video-based frame-level action recognition example from Kinetics dataset~\cite{kay2017kinetics}. Each frame of the video is labeled as to one action or the background~(the background means no interest-of-action in this frame).}
    \label{fig:video}
\end{figure}

\begin{figure}[t]
    \centering
    \includegraphics[width=1\textwidth]{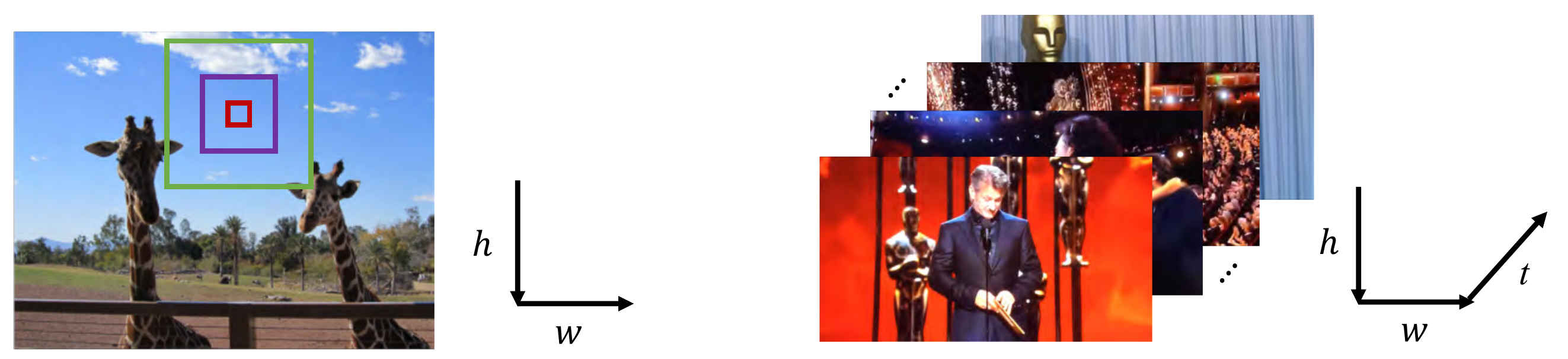}
    \caption{Comparison between the spatial convolutional operation (Left), and the spatio-temporal convolutional operation (Right). }
    \label{fig:st}
\end{figure}

\subsection{Insights from Image-based Semantic Segmentation}

\textbf{Analysis.} As COCO-Stuff dataset~\cite{caesar2018coco} examples shown in Fig.~\ref{fig:image}, the task of image-based semantic segmentation is to generate an object or thing label, such as giraffe, person, boat, sea and sand, for every pixel of images. It is an intuition that we can hardly tell that one blue pixel~(BP) belongs to sky if we only stare at this BP. However if we view more neighboring blue pixels even the white cloud, we then can say this BP must be sky confidently. 
Intuitively as shown in Fig.~\ref{fig:st}~(left), it doesn't make sense to label a blue pixel~(BP) as sky if not taking the neighboring pixels into account. However, when the blue pixel is surrounded by a batch of blue pixels or even white pixels labeled as cloud, it is much more reasonable to regard the BP as part of the sky. The example above illustrates that pursuing larger field of view~(FOV) is critical for semantic segmentation.
In computer vision, spatial convolutional kernels generally sweep along the height and width of one image to generate semantic features of the image, as shown in Fig.~\ref{fig:st}. And to gain larger FOVs, global pooling layers are employed for the whole-image view which advances the high-level understanding on  images~\cite{long2015fully,zhao2017pyramid,chen2018deeplab}. Besides, dilated convolutions~\cite{yu2015multi} enable deep networks to gain multi-scale FOVs and are widely used in semantic segmentation~\cite{lin2017feature,chen2018deeplab,hamaguchi2018effective,lin2017refinenet}. In addition, frameworks that enable combination between low-level features~(small FOV) and high-level features~(large FOV) of images are also popular in semantic segmentation~\cite{ronneberger2015u,lin2017feature,chen2018encoder}.

\textbf{Insight.} Similar to image-based semantic segmentation, in WiSLAR, doing an action usually lasts for a while, thus we should enable our deep network larger and various FOVs over WiFi distortion for the local and global understanding of the distortion.

\subsection{WiSLAR and Video-based Action Recognition}\label{sec:video}

\textbf{Analysis.} As one Kinetics~\cite{kay2017kinetics} example shown in Fig.~\ref{fig:video}~(one awarded Oscar and audiences applauding), video-based action recognition generates one action label or background~(no interest-of-action) for every frame of videos. As a useful practice, stacking multiple continuous frames to form a tensor  with multiple channels can promote the performance of action recognition~\cite{simonyan2014two,karpathy2014large,lan2015beyond}. In the early years, this tensor is considered as one frame with thick channels and swept by 2D spatial convolutional kernels that resemble Fig.~\ref{fig:st}~(left). Then features learned among all stacked frames are severed for video-based action recognition. Recently, 3D convolutional kernels sweeping along the height, width, and channel of stacked tensor is widely applied to improve the performance of action recognition~\cite{ji20133d,hara2018can,carreira2017quo,kay2017kinetics}, illustrated in Fig.~\ref{fig:st}~(right). Compared to 2D spatial convolutional kernels, 3D convolutional kernels can learn extra relations between neighboring frames in multiple scales, which we call time of view~(TOV). Despite 3D convolutional kernels,  Convolution-Deconvolution-Convolution~\cite{shou2017cdc} and Temporal Preservation Convolution~\cite{yang2018exploring} are also proposed to enhance the understanding in TOV for video-based action recognition.  

\textbf{Insight.} In WiSLAR, compared to applying convolutional layers over whole samples like 2D spatial convolutional kernels on all stacked frames in video-based action recognition, we should better apply temporal convolutional kernels in WiFi distortion for multi-scale TOV, which is proven useful for action recognition.

\subsection{Deep Network Design Consideration}\label{sec:proposal}
Based on above two insights, we list three keywords, i.e.,  multi-scale, FOV, and TOV, to highlight the guidelines in network design. More precisely for WiSLAR, the network (1) should take features in small FOVs and large FOVs into account for action recognition, and (2) should apply convolutions to sweep in temporal for multi-scale TOVs. To achieve WiSLAR, we enable Unet~\cite{ronneberger2015u}~(meeting the first requirement), the ability of learning temporal features, which meets the second requirement. The network is extremely simple yet efficient and targets to WiFi-based sample-level action recognition. We call it \textbf{Temporal Unet} and describe it in detail in the following section.

%% file: tex/t-unet.tex
\section{Temporal Unet}\label{sec:tunet}

\begin{figure}[t]
    \centering
    \includegraphics[width=1\textwidth]{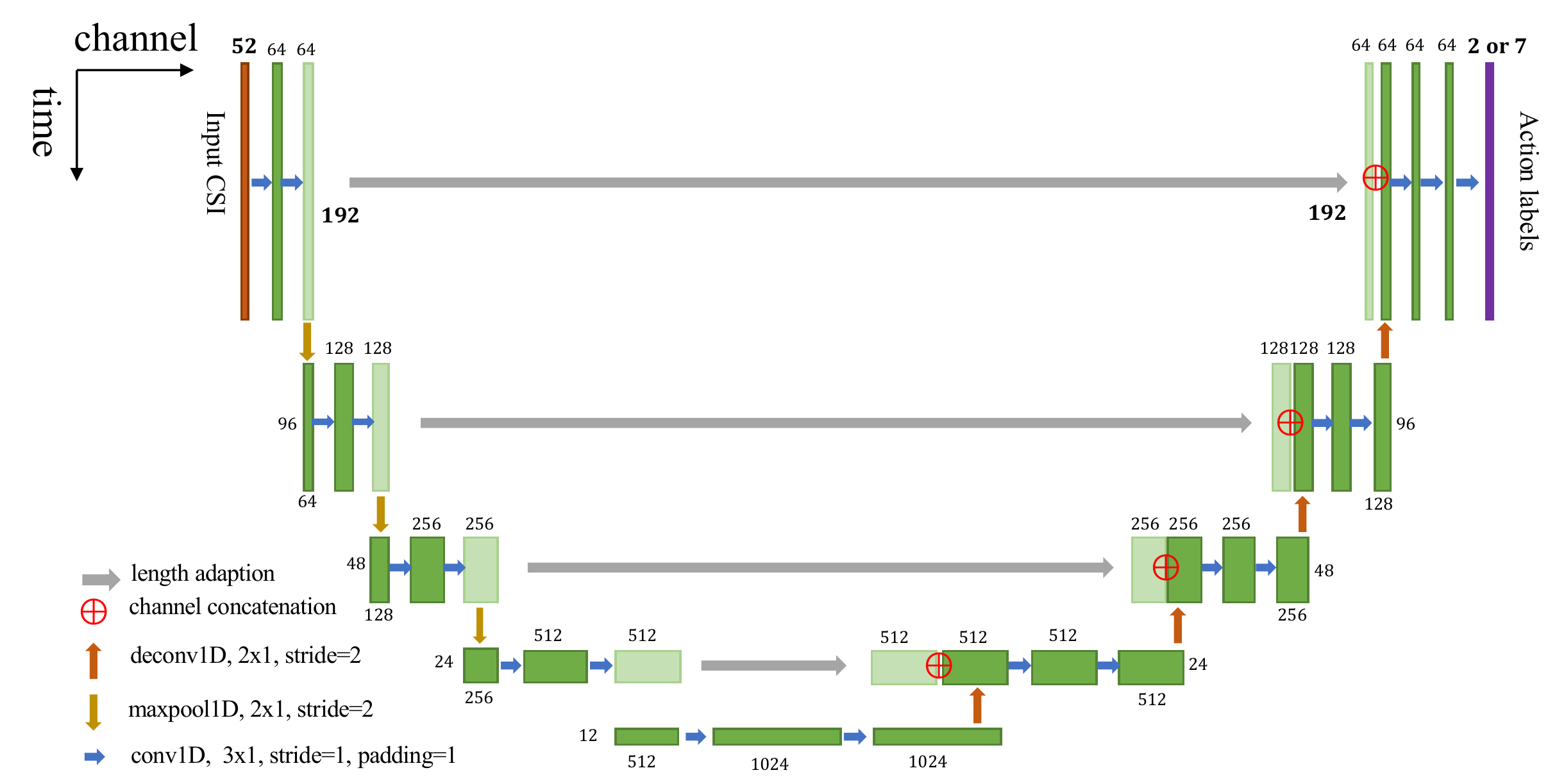}
    \caption{Temporal Unet framework. }
    \label{fig:network}
\end{figure}

\subsection{Overview}
Channel state information~\cite{halperin2011tool} is used as WiFi distortion for sample-level action recognition, which is comprised of the information of all Orthogonal Frequency Division Multiplexing~\cite{nee2000ofdm} carriers between the WiFi sender and the WiFi receiver. We denote one CSI distortion series as $\mathrm{C}=\left \{ c_i; i\in[1,2,...,n] \right \}$, where $c_i$ is the i-th CSI sample, and $n$ is the length of sampled CSI distortion series. Meanwhile we denote the corresponding action label series as $\mathrm{A}=\left \{ a_i; i\in[1,2,...,n] \right \}$, where $a_i$ is the action label of $c_i$. As shown in Fig.~\ref{fig:network}, Temporal Unet~(T-Unet) is to learn a mapping function from CSI sample series to action label series, i.e., $f: \mathrm{C} \rightarrow \mathrm{A}$.

\subsection{Details}

T-Unet contains three main components, i.e., the `down', the `up' and the `shortcut links'. The `down' reduces the size of input CSI series by convolutional layers and max pooling layers that work along the time axis. With layers being deeper and deeper, the `down' can learn action features in larger and larger time of views, which has proven efficient in action recognition as described in Section.~\ref{sec:video}. The `up' increases the feature map size to meet the requirement that the lengths of $A$ and $C$ should be the same. Besides, with more convolutional layers and deconvolutional layers, T-Unet can gain deeper features for WiFi-based sample-level action recognition. Last but not least, the `shortcut links'~(grey arrows in Fig.~\ref{fig:network}) advance feature maps in deeper layers with information directly from lower layers, promoting T-Unet with multi-scale filed of views over the CSI series.

\begin{figure}[t]
    \centering
    \includegraphics[scale=1]{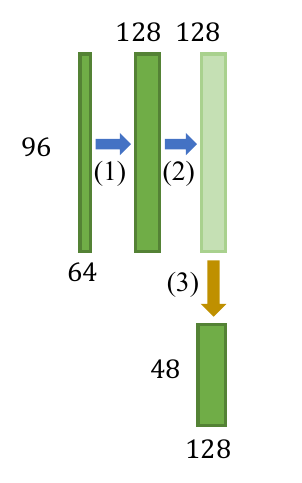}
    \hspace{40pt}
     \includegraphics[scale=1]{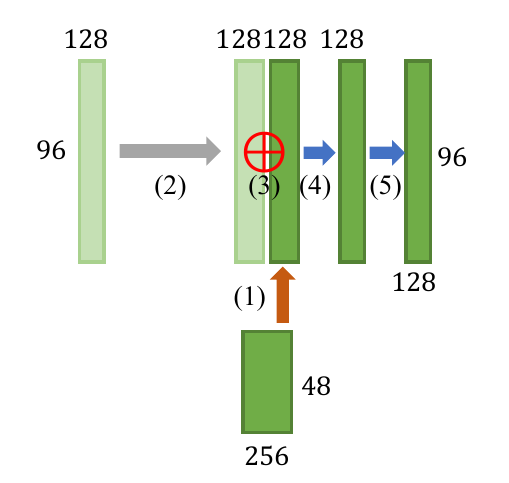}
    \caption{Block examples of T-Unet, i.e.,  `down-96-48'~(left) and  `up-48-96'~(right).}
    \label{fig:block}
\end{figure}

Next we go more details in the main block by specifying network parameter setting. As shown in Fig.~\ref{fig:network}, every two convolutional layers and one max pooling contribute one `down' block, one of which is shown in Fig.~\ref{fig:block}~(left). We name it `down-96-48', where the 96 and 48 are the sizes of input and output in time axis, respectively. In `down-96-48', the input size is $96\times64$, and is converted to feature maps with size of $48\times64$ by two consecutive convolutional layers~(kernel size = $2\times 1$, stride=1, padding=1). Then a max pooling layer~(kernel size=$2\times1$, stride=2, padding=0) makes the converted feature maps $48\times 128$. Then the output of the `down-96-48' is the input of `down-48-24' for further operations. In addition, a `up-48-96' is shown in Fig.~\ref{fig:block}~(right), in which the input~($48\times256$) is up-sampled by a deconvolutional layers~(kernel size =$2\times 1$, stride=2, padding=0) to be $96\times 128$. Meanwhile the `shortcut link' concatenates the up-sampled feature maps with ones in low layer along the channel axis, doubling the channel. Then the doubled feature maps are inputted to two convolutional layers for the output, and the output `up-48-96' is the input of the next `up'. With these three components, T-Unet achieves WiFi-based sample-level action recognition.

\subsection{Loss Function}

We apply Cross-Entropy loss function~\cite{krizhevsky2012imagenet} to optimize T-Unet.  

\textbf{Sample-level action detection.} As shown in Fig.~\ref{fig:first}, sample-level action detection is a binary classification task, 
detecting whether or not one is doing interest-of-action.
predicting whether one does interest-of-action or not at the moment when corresponding sample recorded. In this case, $A$ belongs to $n\times 2$. We compute Cross-Entropy loss between the action label annotations and the predictions.

\textbf{Sample-level action classification.} Also shown in Fig.~\ref{fig:first}, sample-level action classification is a multi-class classification task, classifying what interest-of-action one is doing.
predicting what interest-of-action one does at the moment when corresponding sample recorded. In this case, $A$ belongs to $n\times(cls+1)$, where $cls$ is the number of interest-of-action categories, and $1$ is for the non-action state. Similar to the above, we compute Cross-Entropy loss between the action label annotations and the predictions.

\subsection{Implementation}
We implement T-Unet with PyTorch 1.0.0 with the batch size of 128 and the initial learning rate of 0.005. We train T-Unet with Adam optimizer~\cite{kingma2014adam}~($\beta_1=0.9, \beta_2=0.999$) for 200 epochs, and decrease the learning rate with a decay of 0.5 every 10 epochs. T-Unet is trained in a desktop with one Titan XP GPU. Before each epoch, the training dataset is shuffled.

\

%% file: tex/results.tex
\section{Experiments}\label{results}

\subsection{Dataset} 

We use CSI distortion dataset released in~\cite{wang2019joint}. The dataset contains CSI series that correspond to six gesture actions, i.e, `hand up', `hand down', `hand left', `hand right', `hand circle', and `hand cross'. The action start point and end point of all CSI series are manually annotated. The training set is comprised of 1116 WiFi series, with 6 categories of gestures conducted by one volunteer at 16 indoor locations. The size of test set is 278. The size of one CSI series is $192\times 52$, where the 192 is the number of samples in one series, and the 52 is the number of OFDM data carriers. 

\subsection{Metrics}

We use prediction accuracy over all samples as one metric to evaluate T-Unet on this dataset. Meanwhile we propose the action recognition average precision~(AP) as another metric for more fine-grained evaluation.  

\begin{equation}
    \mathrm{AP}@a = \frac{1}{N} \Sigma_{i=1}^{N} \mathcal{I} (acc_i \geq a), i \in [1,N]
\end{equation}
where $i$ is the index of test series; $N$ is the volume of test set~(in this test set, N=278), $acc_i$ is the sample-level action recognition on the $i$-th CSI series; $\mathcal{I(\cdot)}$ is an indicator that outputs $+1$ if the input is true, whereas outputs 0. The higher of $\mathrm{AP}@a$, the better of T-Unet on the dataset.

\subsection{WiFi-based Sample-level Action Detection}

\begin{figure}[t]
    \centering
    \includegraphics[scale=0.4]{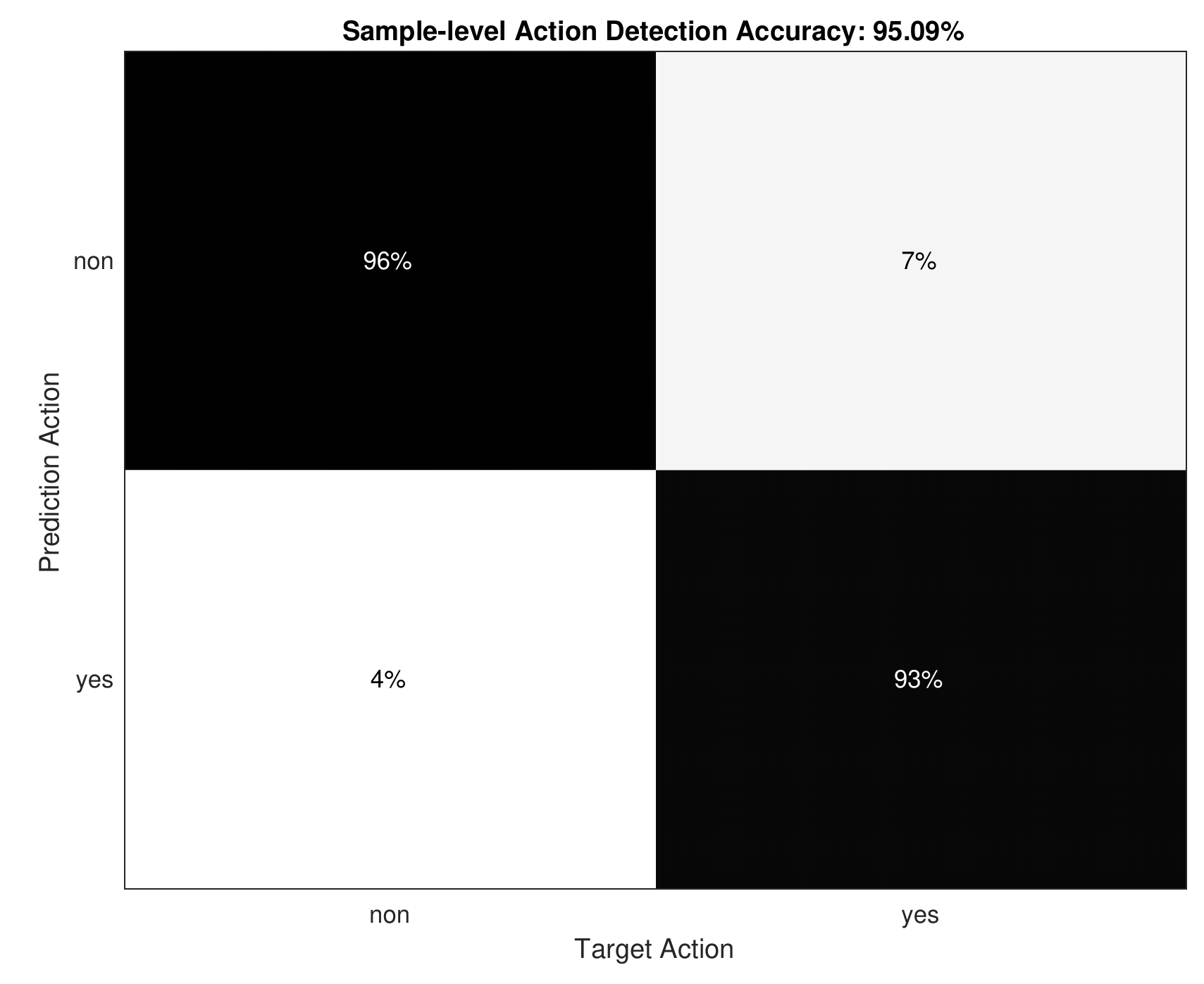}
    \includegraphics[scale=0.4]{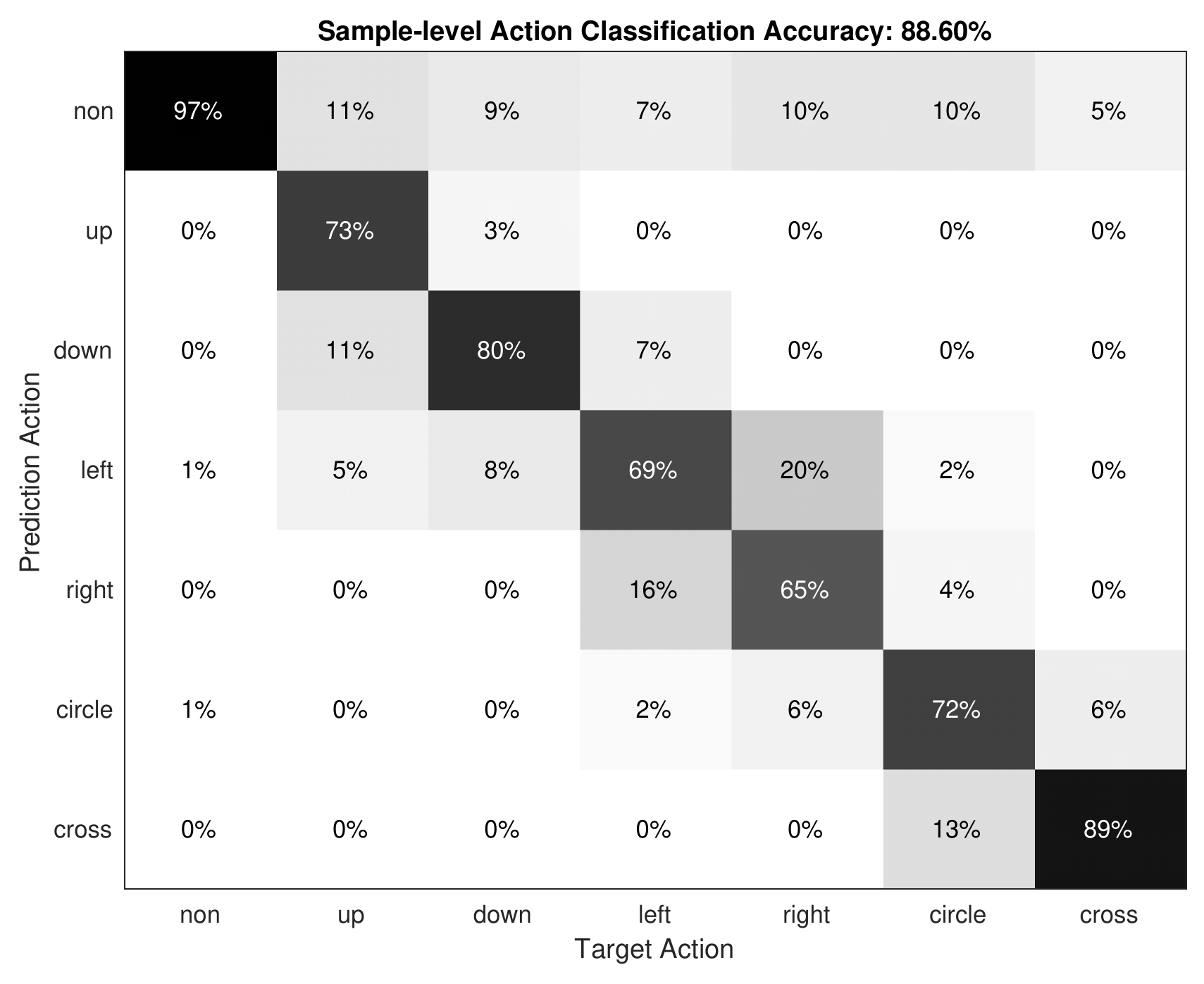}
    \caption{Confusion matrix of T-Unet on~\cite{wang2019joint}.}
    \label{fig:conf}
\end{figure}

\begin{figure}[t]
    \centering
    \includegraphics[width=1\textwidth]{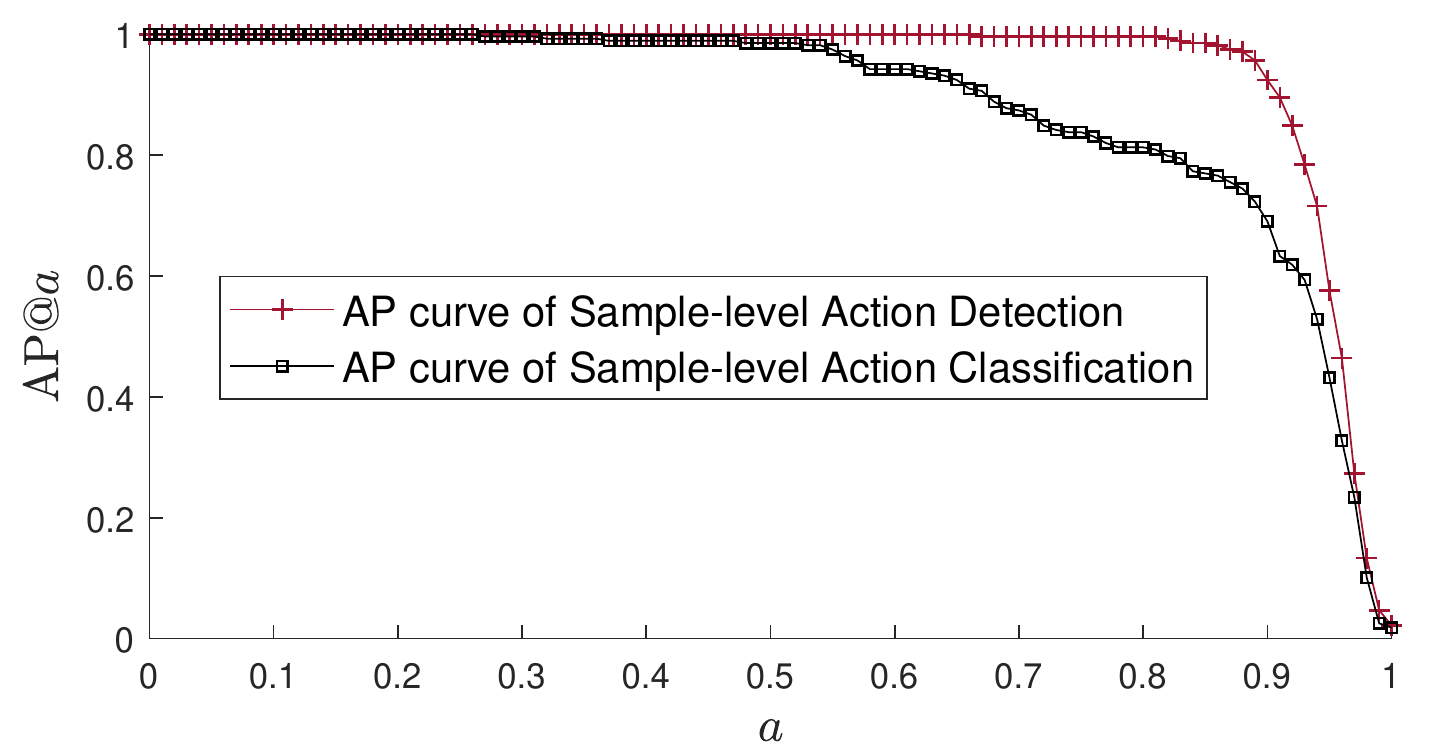}
    \caption{AP curves of T-Unet~\cite{wang2019joint}.}
    \label{fig:ap}
\end{figure}

The overall accuracy of action detection is 95.09\% and the confusion matrix (Fig.~\ref{fig:conf}~left) demonstrates more details about the results. According to the confusion matrix, we see T-Unet works well in distinguishing the states between `non-action' and `doing-action'. In addition, AP curves shown in Fig.~\ref{fig:ap} show performance in more fine-grained view. Taking the point of $(a=0.9, \mathrm{AP}@a=0.92)$ as an example to explain, if we consider it one success when the accuracy is greater than or equals to 0.9, then the success rate of T-Unet on the test set is 0.92, indicating that $ 278\times0.92 = 256$ CSI series are detected successfully. Thus with $a$ increasing, the success rate~($\mathrm{AP}@a$) would decreases. As shown in Fig.~\ref{fig:ap}, $\mathrm{APs}$ of sample-level action detection are with high values until $a=0.9$, which means T-Unet works well even we take 0.9 as the success threshold. Further we average AP@0.5, AP@0.6, AP@0.7, AP@0.8, and AP@0.9, for the mean AP, i.e., 0.98, as one comprehensive metric to assess the performance of T-Unet, which is pretty high.

\begin{table}[ht]
\centering
\caption{APs and mean APs of T-Unet on \cite{wang2019joint}.}
\begin{tabular}{|c|c|c|c|c|c|c|}
\hline
$Task$ & $\mathrm{AP}@0.5$   & $\mathrm{AP}@0.6$   & $\mathrm{AP}@0.7$  & $\mathrm{AP}@0.8$   & $\mathrm{AP}@0.9$   & mean AP  \\ \hline
Action Detection & 1    & 1    & 1    & 1    & 0.92 & 0.98 \\ \hline
Action Classification & 0.99 & 0.94 & 0.87 & 0.81 & 0.69 & 0.86 \\ \hline
\end{tabular}
\label{tab:ap}
\end{table}

\subsection{WiFi-based Sample-level Action Classification}

Similarly, the confusion matrix for action classification is also shown in Fig.~\ref{fig:conf}. We find the major error is wrongly predicting 20\% CSI series of `hand right' to `hand left'. Except some misclassification between `hand right' and `hand left', we notice that classifying other actions as `non-action' state contributes a large part of the misclassification. When looking deep into these wrongly predicted samples, we find that these type of error occurs during the transition series between `non-action' state and `doing-action' state. 
In all, T-Unet achieves an average accuracy of 88.60\% for identifying different categories of actions. In addition, AP curve is plotted in Fig.~\ref{fig:ap}, it gradually decreases when $a=0.5$ and has a sharp decline till $a=0.9$, which indicates that T-Unet performs satisfactorily when the success threshold is not set severe. APs and the mean AP are listed in Table.~\ref{tab:ap}. Compared with quantitative results with WiFi-based action detection, we find WiFi-based action classification is more challenging.

\subsection{Result Visualization}

\begin{figure}[t]
    \centering
  \vspace{10pt}
  
    \hspace{-26pt}
    \includegraphics[width=1.05\textwidth]{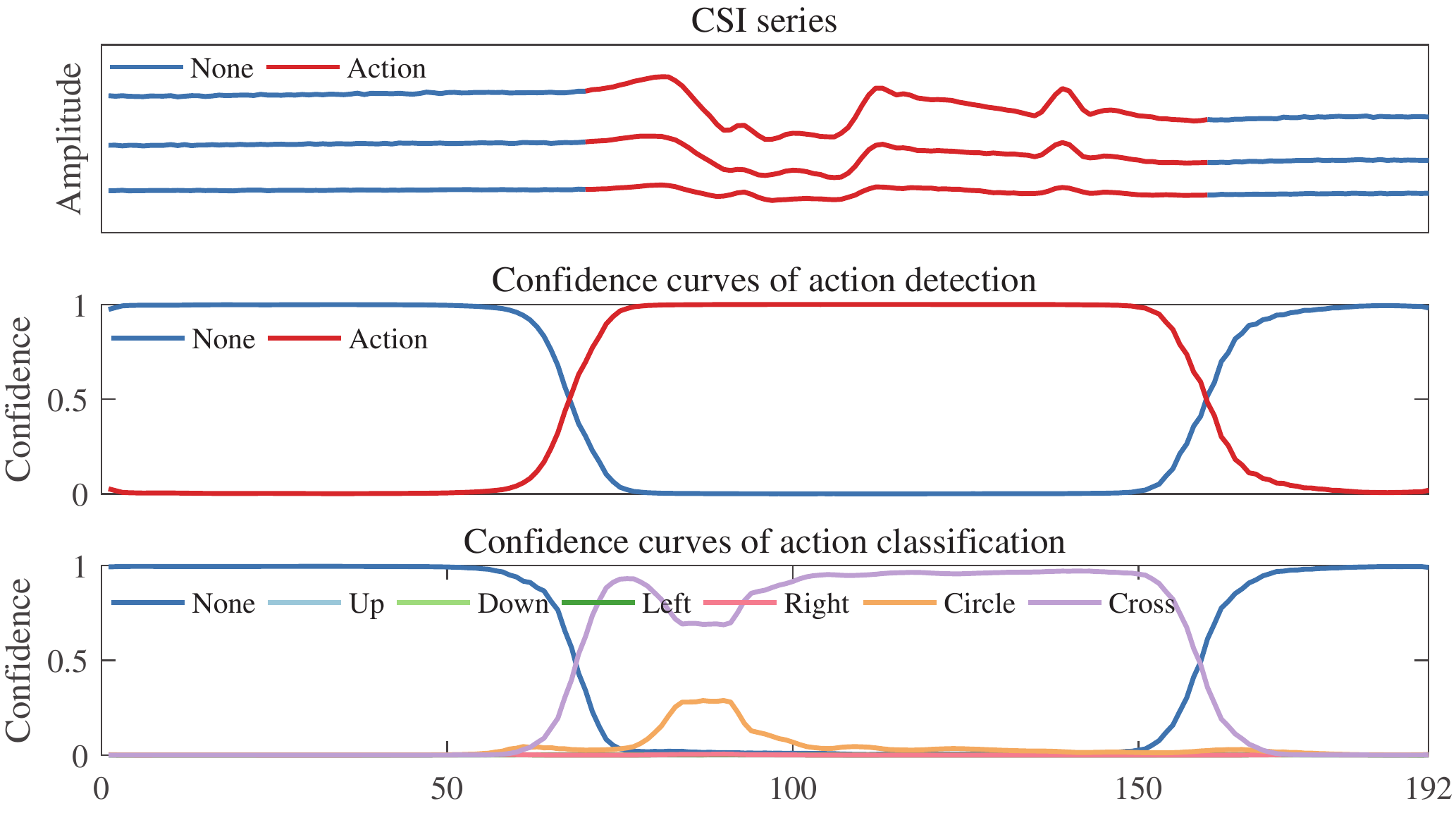}
    \caption{One `hand cross' result example. CSI series vary when one does `hand cross' action~(1st). The confidence curves of sample-level action detection is shown in 2nd sub-figure. Besides the confidence curves of sample-level action classification are shown in the 3rd su-figure.}
    \label{fig:result}
\end{figure}

We present a result example of WiFi-based sample-level action detection and WiFi-based sample-level action detection on `hand cross' in Fig.~\ref{fig:result}. The first sub-figure illustrates time-serial CSI distortion of 3 OFDM carriers~(the 8th, 27th, and 40th carriers), where the blue duration and the red duration are manually labeled as `non-action' and `doing-action', respectively. The blue and the red curves in the middle sub-figure represent sample-level action detection confidence on `non-action' and `doing-action' states, respectively. T-Unet classifies one sample as one specific state that has the highest confidence. Based on this, we can infer that the person first keeps `non-action' state until around 70th~(blues higher than reds), then does one action for time of around 100 samples, and completes the action around 170th sample, then keeps `non-action' state till last. Moreover, the last sub-figure illustrates the confidence curves of action classification for all actions. From it, we can not only infer the moments of action start and end, but also the specific action on every sample. 
Other five action examples on `hand up', `hand down', `hand left', `hand right', and `hand circle' are shown in the Appendix.

%% file: tex/appendix.tex
\section*{Appendix}\label{sec:app}

\begin{figure}[h]
    \centering
    \hspace{-20pt}
    \includegraphics[width=1\textwidth]{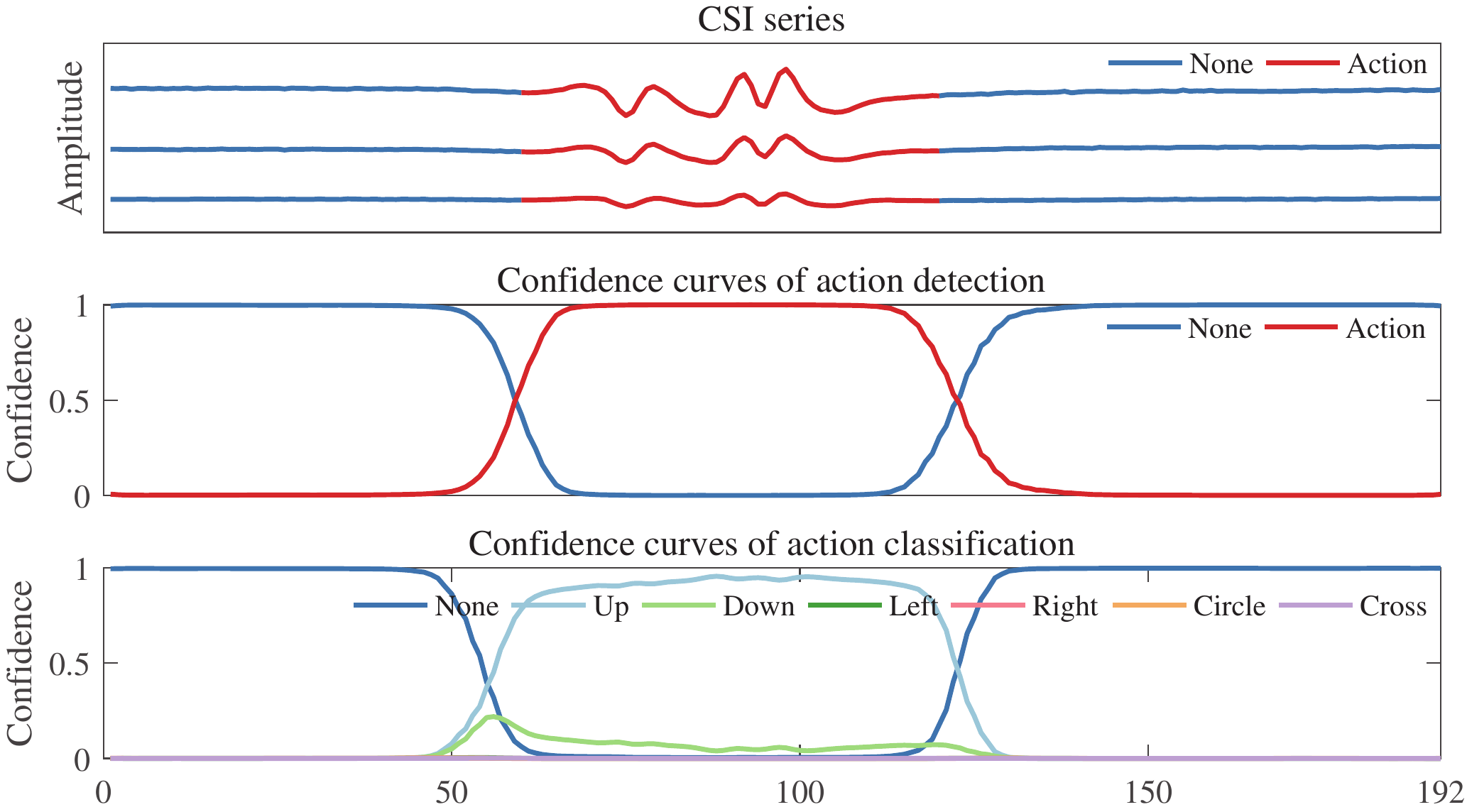}
    \caption{One `hand up' result example. CSI series vary when one does `hand up' action~(1st). The confidence curves of sample-level action detection is shown in 2nd sub-figure. Besides the confidence curves of sample-level action classification are shown in the 3rd su-figure.}
    \label{fig:up}
\end{figure}

\begin{figure}[t]
    \centering
    \hspace{-20pt}
    \includegraphics[width=1\textwidth]{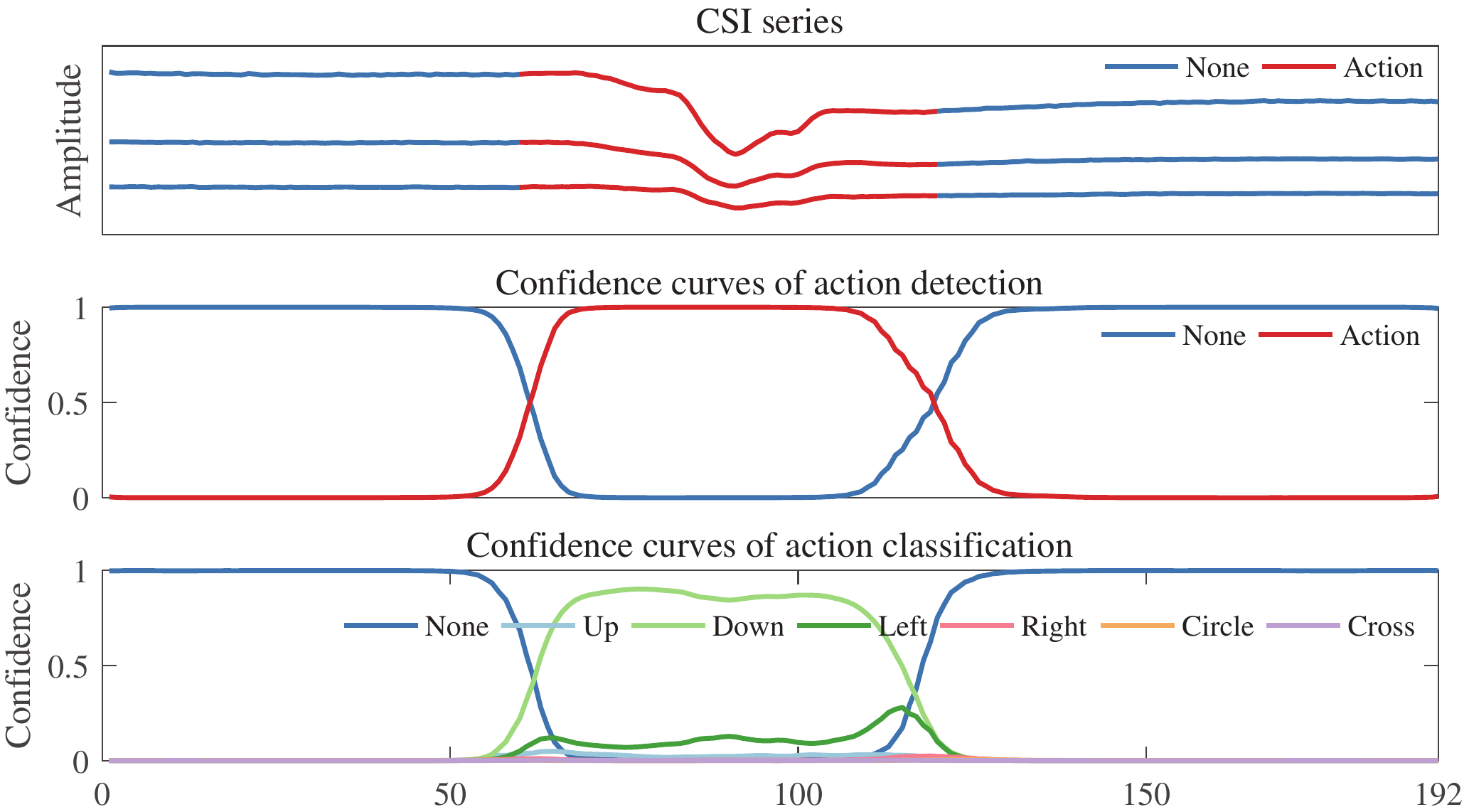}
    \caption{One `hand down' result example. CSI series vary when one does `hand down' action~(1st). The confidence curves of sample-level action detection is shown in 2nd sub-figure. Besides the confidence curves of sample-level action classification are shown in the 3rd su-figure.}
    \label{fig:down}
\end{figure}

\begin{figure}[t]
    \centering
    \hspace{-20pt}
    \includegraphics[width=1\textwidth]{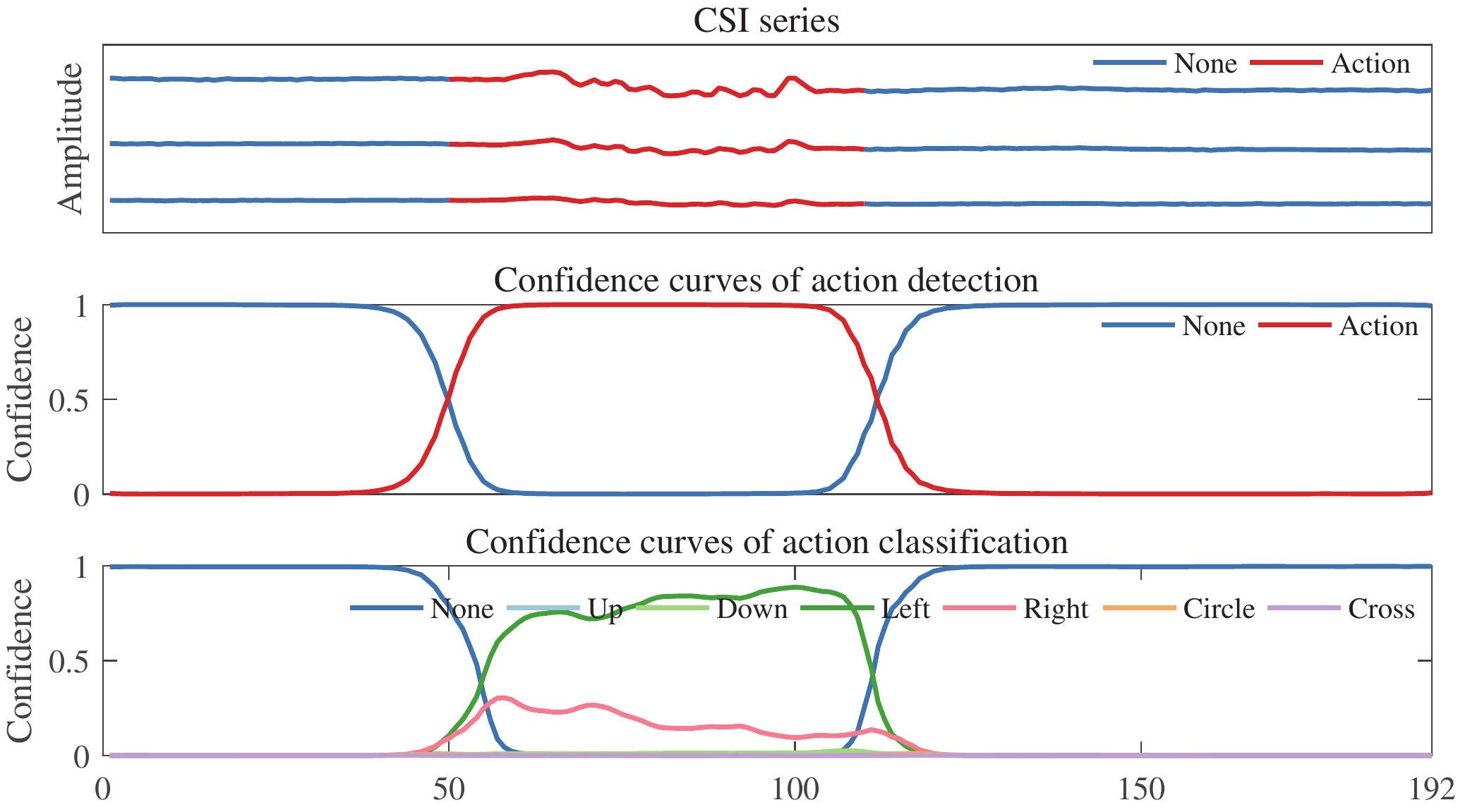}
    \caption{One `hand left' result example. CSI series vary when one does `hand left' action~(1st). The confidence curves of sample-level action detection is shown in 2nd sub-figure. Besides the confidence curves of sample-level action classification are shown in the 3rd su-figure.}
    \label{fig:left}
\end{figure}

\begin{figure}[t]
    \centering
    \hspace{-20pt}
    \includegraphics[width=1\textwidth]{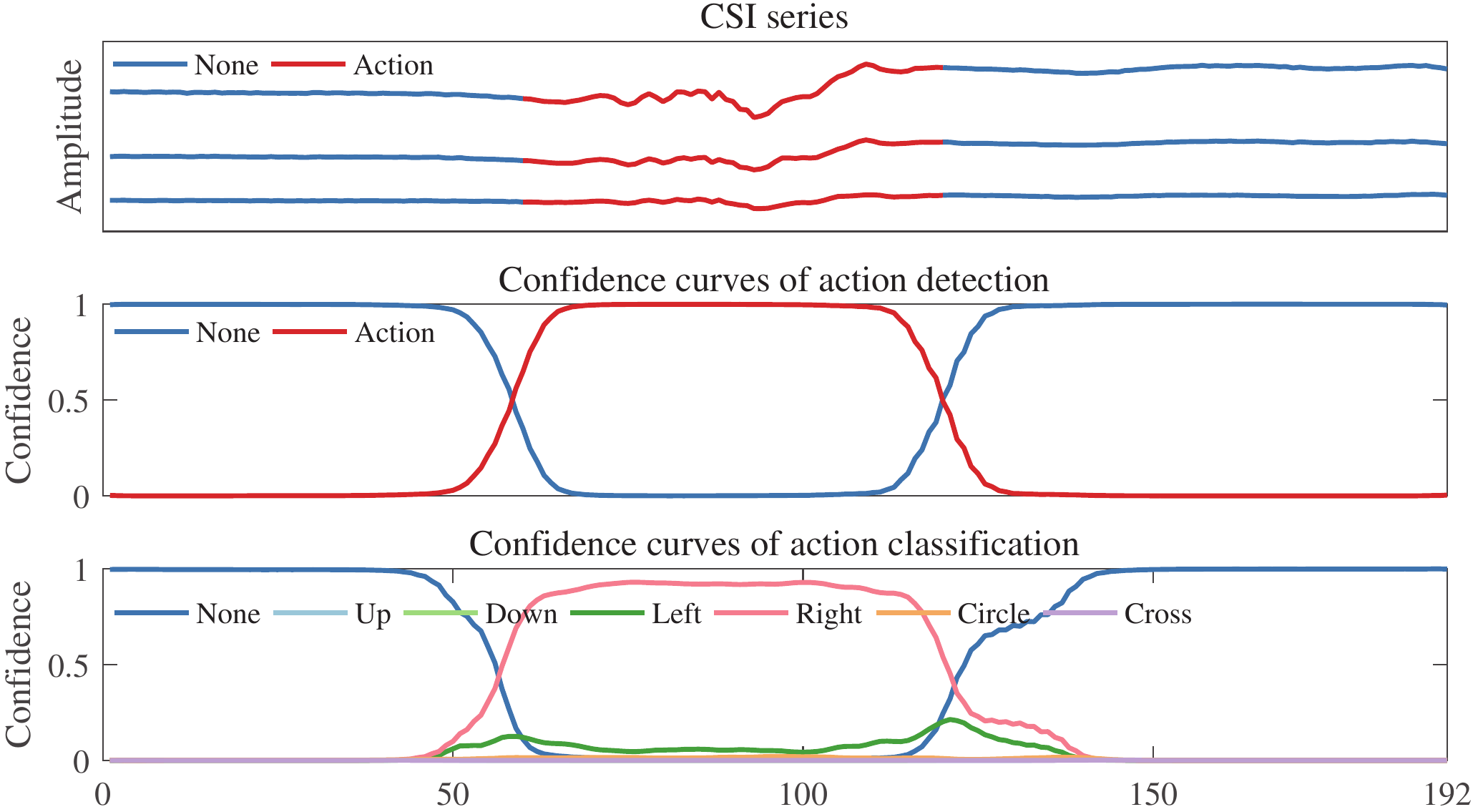}
    \caption{One `hand right' result example. CSI series vary when one does `hand right' action~(1st). The confidence curves of sample-level action detection is shown in 2nd sub-figure. Besides the confidence curves of sample-level action classification are shown in the 3rd su-figure.}
    \label{fig:right}
\end{figure}

\begin{figure}[t]
    \centering
    \hspace{-20pt}
    \includegraphics[width=1\textwidth]{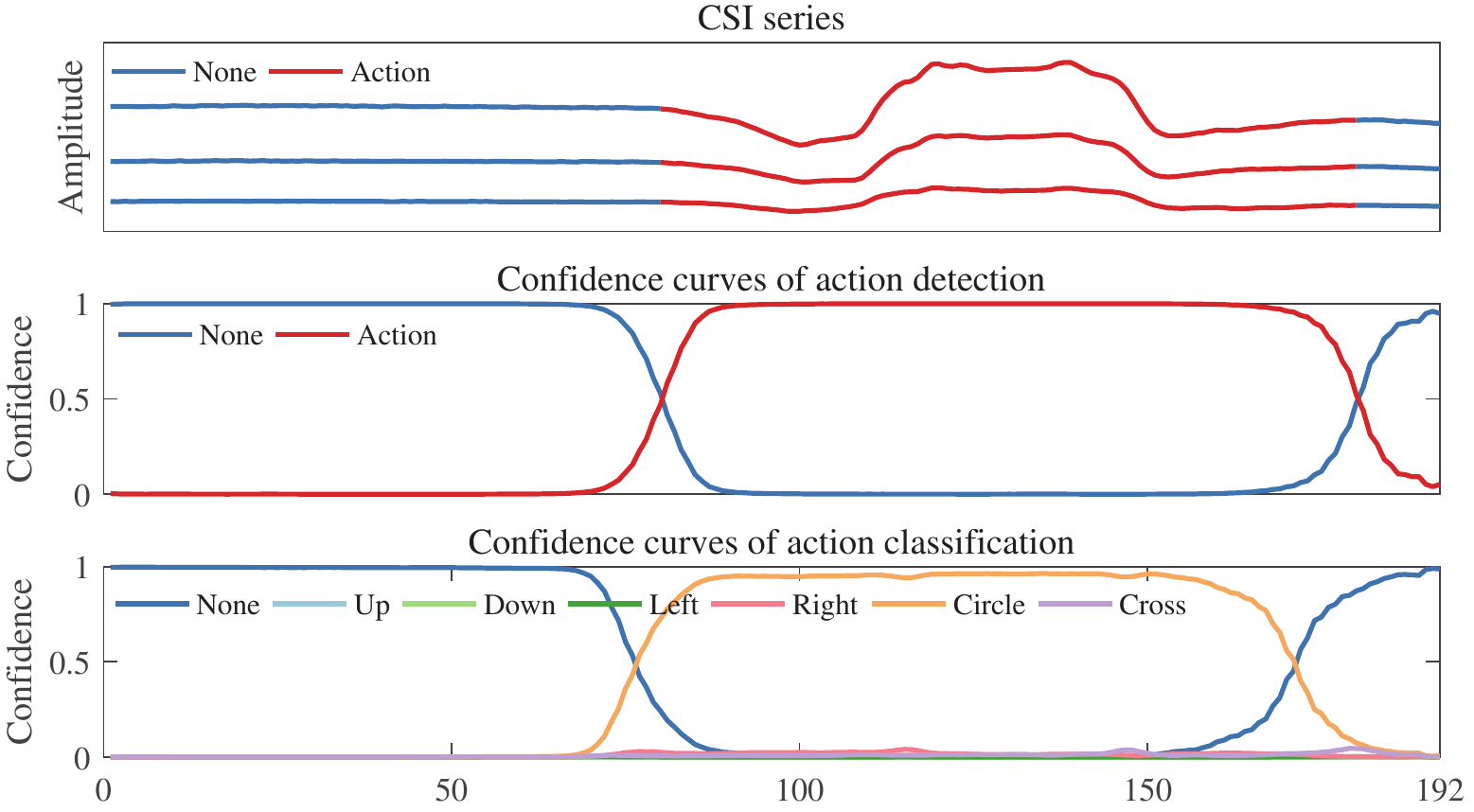}
    \caption{One `hand circle' result example. CSI series vary when one does `hand circle' action~(1st). The confidence curves of sample-level action detection is shown in 2nd sub-figure. Besides the confidence curves of sample-level action classification are shown in the 3rd su-figure.}
    \label{fig:cricle}
\end{figure}